\documentclass[prl,aps,twocolumn,epsfig,superscriptaddress,showpacs]{revtex4}
\usepackage{amsfonts}
\usepackage[dvips]{graphicx}

\usepackage[intlimits]{amsmath}

\begin{document}
\author{Jing-Ling Chen}
\email{chenjl@nankai.edu.cn} \affiliation{ Liuhui Center for Applied
Mathematics and Theoretical Physics Division, Chern Institute of
Mathematics, Nankai University, Tianjin 300071, People's Republic of
China}
\author{Kang Xue}
\affiliation{Department of Physics, Northeast Normal University,
 Changchun, Jilin 130024, People's Republic of China}
\author{Mo-Lin Ge}
\email{geml@nankai.edu.cn} \affiliation{ Liuhui Center for Applied
Mathematics and Theoretical Physics Division, Chern Institute of
Mathematics, Nankai University, Tianjin 300071, People's Republic of
China}

\date{\today}

\title{All Pure Two-Qudit Entangled States Can be Generated via a Universal Yang--Baxter Matrix
Assisted by Local Unitary Transformations}

\begin{abstract}
We show that all pure entangled states of two $d$-dimensional
quantum systems (i.e., two qudits) can be generated from an initial
separable state via a universal Yang--Baxter matrix if one is
assisted by local unitary transformations.
\end{abstract}

  \maketitle

Entangled state is the cornerstone of quantum information theory
\cite{Niels} that has many successful applications in quantum
information processing, such as the revolutionary one-way quantum
computer \cite{Raussendorf}, quantum cryptography \cite{Ekert},
dense coding \cite{Benne1}, teleportation \cite{Benne2},
communication protocols and computation \cite{Prevedel}.
Consequently, the ability to generate and control quantum entangled
states has become a far-reaching goal in experimental manipulation
as well as theoretical investigation in recent years. In fact, a
great number of experiments have been devoted to investigating the
production of entangled states of photons (including the
hyperentangled photon pairs) via the process of spontaneous
parametric down-conversion in a nonlinear optical crystal
\cite{Kwiat1}\cite{Kwiat2}, particularly for use in the tests of
Bell inequalities.

A fundamental notion in quantum computation (QC) is universality: a
set of quantum logic gates (i.e., unitary matrices) is said to be
``universal for QC " if any unitary matrix can be approximated to
arbitrary accuracy by a quantum circuit involving only those gates
\cite{Niels}. For example, an arbitrary $U(2)$ matrix can be
obtained by combining the Hadamard gate together with the phase
shift gates. When such a $U(2)$ matrix is prepared, an arbitrary
state for a single qubit $|\psi\rangle=
\cos\frac{\theta}{2}|0\rangle + \sin\frac{\theta}{2} e^{i \phi}\;
|1\rangle$ can be immediately generated by acting the $U(2)$ matrix
on the initial state $|0\rangle$. Since entangled states are
important for quantum information processing, it gives rise to a
natural question whether an arbitrary pure entangled state of two
qudits (such as the maximally entangled state) can be generated via
a universal matrix from an initial separable state $|00\rangle$?

The purpose of this work is to provide a positive answer to the
above question. We shall study the problem from the theoretical
point of view based on the Yang--Baxter equation, which was
originated in solving quantum integrable models
\cite{YangCN,Baxter}, but recently has been shown to have a deep
connection with topological quantum computation and entanglement
swapping \cite{Kitaev,Freedman,Preskill,Kauffman,Chen2007}. The
Yang-Baxter equation, in principle, can be tested in terms of
quantum optics \cite{Hu2008}. Let us consider a general pure state
of two qudits, whick takes of the following form:
$|\Psi\rangle=\sum_{i,j=0}^{d-1} \mu_{ij} |i\rangle_A |j\rangle_B$,
where $|i\rangle_A$ and $|j\rangle_B$ are the orthonormal bases of
the Hilbert spaces A and B respectively, and $\mu_{ij}$'s are
complex numbers satisfied the normalization condition
$\sum_{i,j=0}^{d-1} |\mu_{ij}|^2=1$. After performing an appropriate
local unitary transformation, the general state $|\Psi\rangle$ can
be recast to a Schmidt-form as $|\psi\rangle_{2-qudit}=\kappa_0
|00\rangle+\kappa_1 |11\rangle+\cdots+\kappa_{d-1} |d-1,
d-1\rangle$,
where $\kappa_j$'s $(j=0,1,...,d-1)$ are the Schmidt coefficients.
Since $|\psi\rangle_{2-qudit}$ is equivalent to $|\Psi\rangle$ up to
a local unitary transformation, hereafter when refer to generating
an arbitrary pure state of two qudits we mean generating the state
$|\psi\rangle_{2-qudit}$. Now we illustrate the problem by starting
from unitary solutions of Yang--Baxter equation in the following.

\emph{Unitary solutions of Yang--Baxter equation.} The Yang--Baxter
equation is given by
\begin{equation}\label{YBEquation}
\breve{R}_{i}(x)\breve{R}_{i+1}(xy)\breve{R}_{i}(y)=\breve{R}_{i+1}(y)\breve{R}_{i}(xy)\breve{R}_{i+1}(x).
\end{equation}
Here the notation $\breve{R}_{i}(x) \equiv \breve{R}_{i,i+1}(x)$ is
used, $\breve{R}_{i,i+1}(x)$ implies $\mathbb{I}_1 \otimes
\mathbb{I}_2 \otimes \mathbb{I}_3 \cdots \otimes
\breve{R}_{i,i+1}(x) \otimes \cdots \otimes \mathbb{I}_n$,
$\mathbb{I}_j$ represents the unit matrix of the $j$-th particle,
and $x=e^{i \theta}$ is a parameter related to the degree of
entanglement. Let the unitary Yang--Baxter $\breve{R}$-matrix for
two qudits be the form
\begin{eqnarray} \label{YBR}
\breve{R}_i(x)&=&F(x)[ {\bf 1}_i + G(x) \;  M_i],
\end{eqnarray}
where $F(x)$ and $G(x)$ are some functions needed to determine later
on, ${\bf 1}_i=\mathbb{I}_i\otimes \mathbb{I}_{i+1}$, and the
Hermitian matrices $M_i$'s (i.e., $M_i=M_i^\dag$) satisfy the Hecke
algebraic relations: $(M_i M_{i+1} M_{i}-M_{i+1} M_{i} M_{i+1})+
g(M_i-M_{i+1})=0$, $M_i^2=\alpha M_i +\beta \; {\bf 1}_i$,
with $\alpha=d-2$ and $\beta=g=d-1$. Substituting Eq. (\ref{YBR})
into Eq. (\ref{YBEquation}), one has $G(x)+G(y)+ \alpha G(x) G(y)
=[1+ g G(x) G(y)] \; G(xy)$.
The unitary condition
$\breve{R}^\dag_i(x)=\breve{R}^{-1}_i(x)=\breve{R}_i(x^{-1})$ yields
$G(x)+G(x^{-1})+ \alpha\; G(x) G(x^{-1})=0$, $F(x) F(x^{-1}) [1+
\beta\; G(x) G(x^{-1})]=0$.
In addition, the initial condition $\breve{R}_i(x)=I_i$ leads to
$G(x=1)=0$,  $F(x=1)=1$.
As a result, one has
\begin{eqnarray*}
 G(x)=- \frac{x-x^{-1}}{(d-1)x+x^{-1}}, \;\;
 F(x)= \frac{(d-1)x+x^{-1}}{d}.
\end{eqnarray*}
In this work, the $d^2 \times d^2$ matrix $M$ is realized as
\begin{eqnarray}
M
&=& \sum_{r=1}^{d-1} P_r \otimes P_r = \sum_{i, j=0}^{d-1}
\sum_{r=1}^{d-1} \; |i j\rangle\langle \overline{i+r},
\overline{j+r}|.
\end{eqnarray}
where $\overline{i+r} ={\rm Mod}\;[i+r, d]$, and
\begin{eqnarray}
&&P_r=\sum_{i=0}^{d-1} |i\rangle\langle \overline{i+r}|, \; r=0, 1,
\cdots, d-1,
\end{eqnarray}
are the circulation matrices that transform the basis $\{|r\rangle,
|r+1\rangle, |r+2\rangle, ..., |2\rangle, |1\rangle, |0\rangle\}$ of
a qudit to the basis $\{|0\rangle, |1\rangle, |2\rangle, ...,
|d-3\rangle, |d-2\rangle, |d-1\rangle\}$. The operator $P_r$ can be
realized through the multiplication of permutation operators
$\mathbb{P}_{k, k+1}=(\mathbb{I}-|k\rangle\langle
k|-|k+1\rangle\langle k+1|)+|k\rangle\langle k+1|+|k+1\rangle\langle
k|$ of a single qudit, for example, $P_1=\mathbb{P}_{d-2,
d-1}...\mathbb{P}_{2,3}\mathbb{P}_{1,2}$. Moreover, the traceless
matrices $P_r$'s satisfy the following interesting relations:
$$P_r=(P_1)^r, \;P_m P_n=P_n P_m=P_{{\rm Mod}[m+n, d]}.$$
Let $P_0=\sum_{i=0}^{d-1} |i\rangle\langle i|$ denote the $d \times
d$ unit matrix, we eventually arrive at the unitary Yang--Baxter
matrix for two qudits as
\begin{eqnarray}
\breve{R}_i(x)&=& \frac{1}{d}\biggr\{[(d-1) x+x^{-1}\;] P_0\otimes
P_0\nonumber\\
&&\;\;\;\;-(x-x^{-1})\; \sum_{r=1}^{d-1}P_r\otimes P_r\biggr\},
\end{eqnarray}
which has not been reported in the literature. Our main result of
connecting unitary Yang--Baxter matrices with entangled states of
two qudits is the following Theorem.

\emph{Theorem}: All pure two-qudit entangled states
$|\psi\rangle_{2-qudit}$ can be generated from an initial separable
state $|00\rangle$ via a universal Yang--Baxter matrix
$\breve{R}(x)$ if one is assisted by local unitary transformations
$U_A \otimes U_B$ and $V_A \otimes V_B$, namely,
\begin{equation}\label{YBstate}
|\psi\rangle_{2-qudit}=[V_A \otimes V_B]\; \breve{R}(x) \;[U_A
\otimes U_B]\; |00\rangle.
\end{equation}
Here the local unitary transformation $V_A \otimes V_B$ is
introduced in order to transform a two-qudit state into its
Schmidt-form.

\emph{Proof}. We would like to provide analytical proof for the case
with $d=2$ and numerical proof for the cases with $d=3$ and 4.

i) For $d=2$, in this case $P_1=\left(
\begin{matrix}
0 & 1 \\
1 & 0
\end{matrix}\right).
$ When one acts $\breve{R}(x)$ directly on the separable state
$|00\rangle$, he then generates the following family of states
\begin{equation}\label{YBstate0}
|\psi\rangle_{YB}=\frac{1}{2} \biggr[
(x+x^{-1})|00\rangle-(x-x^{-1})|11\rangle\biggr].
\end{equation}
In Ref. \cite{Fei}, the generalized concurrence (or the degree of
entanglement \cite{Wootters}) for two qudits is given by
\begin{eqnarray}
&&{\cal C}=\sqrt{\frac{d}{d-1}\biggr( 1-I_1\biggr)},
\end{eqnarray}
where $I_1={\rm Tr}[\rho_A^2]={\rm
Tr}[\rho_B^2]=|\kappa_0|^4+|\kappa_1|^4+\cdots+|\kappa_{d-1}|^4$,
$\rho_A$ and $\rho_B$ are the reduced density matrices for the
subsystems. For $d=2$, one easily has ${\cal C}=2|\kappa_0
\kappa_1|$. Obviously, $|\psi\rangle_{YB}$ has already been in the
form of $|\psi\rangle_{2-qubit}=\kappa_0 |00\rangle +
\kappa_1|11\rangle$, with $\kappa_0=(x+x^{-1})/2=\cos\theta$ and
$\kappa_1=-(x-x^{-1})/2=-i\sin\theta$. The degree of entanglement
for the state $|\psi\rangle_{YB}$ equals to ${\cal C}=2|\kappa_0
\kappa_1|=|\sin(2\theta)|$, which may range from 0 to 1. Thus, for
the case of two qubits, all pure states can be generated from
$|00\rangle$ directly via a universal Yang--Baxter matrix
$\breve{R}(x)$. By the way, when $\theta=\pi/4$, the state
$|\psi\rangle_{YB}$ becomes the maximally entangled state, or the
Bell state $|\psi\rangle_{Bell}=(1/\sqrt{2}) (|00\rangle-i
|11\rangle)$.

ii) For $d=3$,  in this case
\begin{eqnarray*} &&P_1=\left(
\begin{matrix}
0 & 1 & 0\\
0 & 0 & 1\\
1 & 0 & 0
\end{matrix}\right), \;\;
P_2=\left(
\begin{matrix}
0 & 0 & 1\\
1 & 0 & 0\\
0 & 1 & 0
\end{matrix}\right).
\end{eqnarray*}
When matrix $\breve{R}(x)$ is acted directly on $|00\rangle$, it
yields the following family of states
\begin{eqnarray}\label{YBstate1}
|\psi\rangle_{YB}=\frac{1}{3} [ (2 x+x^{-1})|00\rangle-(x-x^{-1})
(|11\rangle+|22\rangle)],
\end{eqnarray}
whose generalized concurrence reads
\begin{eqnarray*}\label{doe}
 {\cal C}=\sqrt{\frac{3}{2}\biggr[1-\frac{1}{81}|2
x+x^{-1}|^4-\frac{2}{81}|x-x^{-1}|^4 \biggr]}.
\end{eqnarray*}
When $|2 x+x^{-1}|=|x-x^{-1}|$, namely $x=e^{i\pi/3}$, the state
$|\psi\rangle_{YB}$ becomes the maximally entangled state (here we
would like to call it as the GHZ state) of two qutrits as
$|\psi\rangle_{GHZ}=\frac{-i}{\sqrt{3}} [ \omega|00\rangle+
|11\rangle+|22\rangle]$. In general, if one acts the unitary
Yang--Baxter matrix $\breve{R}(x=e^{i\pi/3})$ on the basis
$\{|00\rangle, |01\rangle, |02\rangle, |10\rangle, |11\rangle,
|12\rangle, |20\rangle, |21\rangle, |22\rangle\}$, he will generate
nine complete and orthogonal maximally entangled states of two
qutrits.

It is easy to check that the generalized concurrence ${\cal C}$
ranges from 0 to 1 when the parameter $\theta$ runs from 0 to $\pi$.
However, this fact does not mean that $|\psi\rangle_{YB}$ is an
arbitrary state of two qutrits, because $|\psi\rangle_{2-qutrit}$
has at least two free parameters while $|\psi\rangle_{YB}$ contains
only one. Actually, the entanglement property of a two-qutrit system
is completely charecterized by two entanglement invariants $I_1={\rm
Tr}[\rho_A^2]={\rm
Tr}[\rho_B^2]=|\kappa_0|^4+|\kappa_1|^4+|\kappa_{2}|^4$ and
$I_2={\rm Tr}[\rho_A^3]={\rm
Tr}[\rho_B^3]=|\kappa_0|^6+|\kappa_1|^6+|\kappa_{2}|^6$, or
equivalently,
\begin{eqnarray}\label{invariant}
&& I'_1=\frac{3}{2}(1-I_1),\nonumber\\
&&I'_2=\frac{9}{8}(1-I_2),
\end{eqnarray}
where the normalized entanglement invariants $I'_1, I'_2 \in [0,
1]$.

\begin{figure}\label{fig1}
\includegraphics[width=85mm]{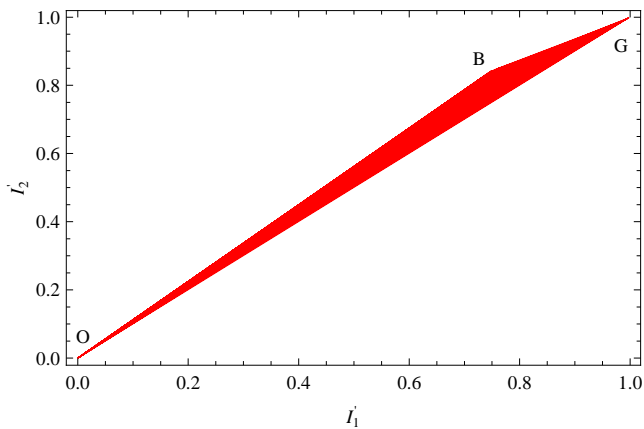}\\
 \caption{(Color online) In the $I'_1-I'_2 $ coordinate,
the separable states,
 such as $|00\rangle$, locates at the origin $O=(0, 0)$; the maximally entangled
 state locates at the point $G=(1,
 1)$, which is the point farthest from the origin; and the entangled
 state,
such as  $\frac{1}{\sqrt{2}} ( |00\rangle+ |11\rangle)$, locates at
the point $B=(\frac{3}{4}, \frac{27}{32})$. One may plot the points
$(I'_1, I'_2)$ for the states $|\psi\rangle'_{YB}=\breve{R}(x)\;
[|0\rangle_A
\otimes(\cos\varphi|0\rangle_B+\sin\varphi|1\rangle_B)]=\breve{R}(x)
\;[U_A \otimes U_B]\; |00\rangle$, and they perfectly recover all
the red region of figure.}
\end{figure}

In Fig.1, we have plots points $(I'_1, I'_2)$ for the two-qutrit
state $|\psi\rangle_{2-qutrit}=\kappa_0 |00\rangle+\kappa_1
|11\rangle+\kappa_{2} |22\rangle$ by randomly taking $10^7$ values
of $\kappa_0$, $\kappa_1$, and $\kappa_2$, see the red region of
figure, whose contour lines form a curved triangle $\Delta OBG$. One
may observe that for a fixed value of $I'_1$, there are different
values for $I'_2$, therefore $I'_1$ is not enough for characterizing
the entanglement property of two qutrits. In the $I'_1-I'_2 $
coordinate, the separable states,
 such as $|00\rangle$, locate at the origin $O=(0, 0)$. The
 maximally entangled states (or say the GHZ states), such as $|\psi\rangle_{GHZ}=\frac{1}{\sqrt{3}} ( |00\rangle+
|11\rangle+|22\rangle)$,
locate at the point $G=(1, 1)$. And the entangled states, such as
$\frac{1}{\sqrt{2}} ( |00\rangle+ |11\rangle)$, locate at the point
$B=(\frac{3}{4}, \frac{27}{32})$. The contour line OB corresponds to
the states $|\psi\rangle_{2-qutrit}^1=\cos\xi |00\rangle+\sin\xi
|11\rangle$, the point $(I'_1, I'_2)$ runs from O to B when $\xi$
runs from 0 to $\pi/2$; The contour lines OG and GB correspond to
the states $|\psi\rangle_{2-qutrit}^2=\cos\xi |00\rangle+\sin\xi
(|11\rangle+|22\rangle)/\sqrt{2}$, the point $(I'_1, I'_2)$ runs
from O to G when $\xi$ runs from 0 to $\pi/3$, and from G to B when
$\xi$ runs from $\pi/3$ to $\pi/2$.

The state $|\psi\rangle_{YB}$ is one part of the state
$|\psi\rangle_{2-qutrit}^2$, when $\theta$ runs from 0 to $\pi/2$,
the point $(I'_1, I'_2)$ runs from O to G, then runs along the line
GB towards to point B and finally stops at a point $(\frac{8}{9},
\frac{25}{27})$, which corresponds to the state
$|\psi\rangle_{YB}=\frac{i}{3} ( |00\rangle-2|11\rangle
-2|22\rangle)$. Namely, when one acts $\breve{R}(x)$ directly on the
state $|0\rangle_A |0\rangle_B$, he cannot get all pure state of two
qutrits. However, numerical computation shows that if one acts
$\breve{R}(x)$ on the state $|0\rangle_A
\otimes(\cos\varphi|0\rangle_B+\sin\varphi|1\rangle_B)$, he can
indeed obtain all pure two-qutrit states: by randomly taking $10^7$
values of $\theta$ and $\varphi$, one may plot points $(I'_1, I'_2)$
for the states $$|\psi\rangle'_{YB}=\breve{R}(x)\; [|0\rangle_A
\otimes(\cos\varphi|0\rangle_B+\sin\varphi|1\rangle_B)],$$ which
perfectly recover all the red region of Fig. 1. This means that if
one is assisted by the local unitary transformations $U_A \otimes
U_B$ and $V_A \otimes V_B$, with $U_A=\mathbb{I}$ and
$U_B=\cos\varphi (|0\rangle \langle 0| - |1\rangle \langle
1|)+\sin\varphi (|0\rangle \langle 1|+ |1\rangle \langle
0|)+|2\rangle \langle 2|$, he then can generated all pure two-qutrit
entangled states in the following way: $|\psi\rangle_{2-qutrit}=[V_A
\otimes V_B]\; \breve{R}(x) \;[U_A \otimes U_B]\; |00\rangle$. This
ends the numerical proof for the case with $d=3$.

\begin{figure}\label{fig2}
\includegraphics[width=75mm]{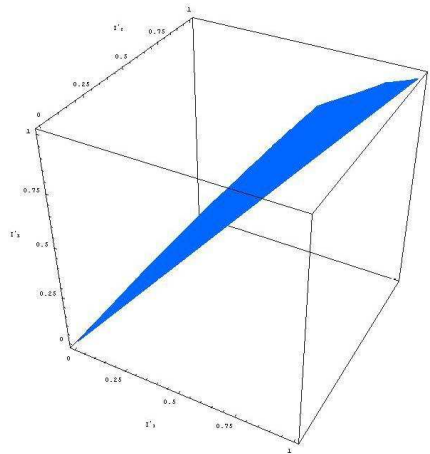}\\
 \caption{(Color online) In the $I'_1-I'_2-I'_3 $ coordinate,
we have plots points $(I'_1, I'_2, I'_3)$ of the two-qudit state
$|\psi\rangle_{2-qudit}$ for $d=4$, see the blue region. By randomly
taking $10^7$ values of $\theta$, $\varphi_1$ and $\varphi_2$, one
may also plot points $(I'_1, I'_2, I'_3)$ for the states
$|\psi\rangle'_{YB}=\breve{R}(x)\; |\Phi\rangle_A \otimes
|\Phi\rangle_B=\breve{R}(x)\;|0\rangle_A \otimes(\cos\varphi_1
|0\rangle_B+\sin\varphi_1 \cos\varphi_2|1\rangle_B++\sin\varphi_1
\sin\varphi_2|2\rangle_B)$,
 which perfectly recover all the blue region of figure.}
\end{figure}

iii) For $d \ge 4$, there are $d-1$ normalized entanglement
invariants for the two-qudit state $|\psi\rangle_{2-qutrit}$, i.e.,
$I'_j=\frac{d^j}{d^j-1}(1-I_j)$, with $I_j={\rm
Tr}[\rho_A^{j+1}]={\rm Tr}[\rho_B^{j+1}]=\sum_{r=0}^{d-1}
|\kappa_r|^{2(j+1)}$, $(j=1,2,...,d-1)$. When acting on the
separable state $|00\rangle$, the unitary Yang--Baxter matrix
$\breve{R}(x)$ generates the following family of states
\begin{eqnarray*}
|\psi\rangle_{YB}=\frac{1}{d} \{[(d-1)
x+x^{-1}]|00\rangle-(x-x^{-1}) \sum_{j=1}^{d-1}|jj\rangle\}.
\end{eqnarray*}
If one requires the state $|\psi\rangle_{YB}$ to be the maximally
entangled state (or the GHZ state), he must set $|(d-1)
x+x^{-1}|=|x-x^{-1}|,$
namely,
\begin{eqnarray*}
\cos(2\theta)=1-\frac{d}{2},
\end{eqnarray*}
For $d=2$, $d=3$,  and $d=4$, one has $\theta=\pi/4$, $\pi/3$, and
$\pi/2$, respectively.
However, the above condition is not valid for $d \ge 5$, because one
will have $|\cos(2\theta)|> 1$ when $d\ge 5$. This fact implies that
the maximally entangled two-qudit states can be generated when we
act $\breve{R}(x)$ directly on the separable state $|00\rangle$ for
$d \le 4$.

Similarly, numerical results show that all pure two-qudit entangled
states can be generated in the following way:
$|\psi\rangle_{2-qudit}=[V_A \otimes V_B]\; \breve{R}(x) \;[U_A
\otimes U_B]\; |00\rangle =[V_A \otimes V_B]\; \breve{R}(x) \;
|\Phi\rangle_A \otimes |\Phi\rangle_B$, where $|\Phi\rangle_A=U_A
|0\rangle_A=|0\rangle_A$ and $|\Phi\rangle_B=U_B
|0\rangle_B=\cos\varphi_1 |0\rangle_B+\sin\varphi_1
\cos\varphi_2|1\rangle_B+\cdots +\sin\varphi_1 \sin\varphi_2\cdots
\sin\varphi_{d-2}|d-2\rangle_B$. In particular, for $d=4$, one has
$|\Phi\rangle_B=\cos\varphi_1 |0\rangle_B+\sin\varphi_1
\cos\varphi_2|1\rangle_B++\sin\varphi_1 \sin\varphi_2|2\rangle_B$.
Numerical proof of the Theorem for $d=4$ is provided in Fig. 2. In
Fig. 2, we have plots points $(I'_1, I'_2, I'_3)$ for the two-qudit
state $|\psi\rangle_{2-qudit}=\kappa_0 |00\rangle+\kappa_1
|11\rangle+\kappa_{2} |22\rangle++\kappa_{3} |33\rangle$ by randomly
taking $10^7$ values of $\kappa_j$'s, see the blue region of figure.
By randomly taking $10^7$ values of $\theta$, $\varphi_1$ and
$\varphi_2$, one may also plot points $(I'_1, I'_2, I'_3)$ for the
states
\begin{eqnarray}\label{YBstate2}
|\psi\rangle'_{YB}=\breve{R}(x)\; |\Phi\rangle_A \otimes
|\Phi\rangle_B,
\end{eqnarray}
 which perfectly recover all the blue region of Fig. 2.

In conclusion, we have shown that all pure entangled states of two
qudits can be generated from an initial separable state $|00\rangle$
via a universal Yang--Baxter matrix if one is assisted by local
unitary transformations. Eventually, we would like to point out that
the spirit of a unitary matrix assisted by local unitary
transformations as shown in Eq. (\ref{YBstate}) or Eq.
(\ref{YBstate2}) coincides with the spirit of \emph{entangling
power}, which is a quantitative measure how much entanglement
capability a given unitary operator has in the context of quantum
information. The concept of entangling power is first introduced in
Refs. \cite{Zan00,Zan01}, which is defined as
\begin{equation}
e_{\text{p}}({\cal U}):=\overline{E({\cal U} \;|\Phi\rangle_A
\otimes |\Phi\rangle_B)}.
\end{equation}
where $E(|\Psi\rangle)=1-{\rm Tr}[\rho_A^2]$, the overbar stands for
the average over all the product states, and it can be simplified as
$e_{\text{p}}({\cal U})=(d/d+1)^{2}\left[ E({\cal U})+E( {\cal U}
{\cal S})-E({\cal S})\right]$, with ${\cal
S}=\sum_{i,j=0}^{d-1}|ij\rangle\langle ij|$ is the permutation
operator of two qudits. The entangling power has been useful for the
study of quantum evolutions and Hamiltonians
\cite{Zan00,Zan01,Dur,Lei02,Wan02,Chi02,Kra01}, and been also
applied to some quantum chaotic systems
\cite{Chao1,Baker1,Baker2,Baker3}. Actually, $E(|\Psi\rangle)=1-{\rm
Tr}[\rho_A^2]$ is  the entanglement invariant $I'_1$ up to a
normalized constant $d/(d-1)$. Similarly, based on the entanglement
invariants of two qudits one may define a series of entangling
powers as $e^j_{\text{p}}({\cal U}):=\overline{E_j({\cal U}
\;|\Phi\rangle_A \otimes |\Phi\rangle_B)}$ with $E_j=I'_j$, which we
will investigate subsequently.

This work is supported in part by NSF of China (Grants No. 10575053
and No. 10605013), Program for New Century Excellent Talents in
University, and The Project-sponsored by SRF for ROCS, SEM.

\end{document}